\documentclass[prl,letterpaper,twocolumn,showpacs,showkeys,lengthcheck,
               floatfix,nofootinbib,preprintnumbers,superscriptaddress]{revtex4-1} 

\usepackage{amsmath,amssymb,amsfonts,graphicx}
\usepackage{bm}
\usepackage{hyperref}
\usepackage{slashed}
\usepackage[caption=false]{subfig}
\usepackage[svgnames]{xcolor}
\usepackage[english]{babel}
\usepackage{blindtext}
\usepackage{microtype}
\usepackage{tikz}

\RequirePackage{slashed}

\def\a{\alpha}
\def\g{\gamma}

\def\s{\sigma}
\def\o{\omega}

\def\hs{\hspace}

\def\lra{\longrightarrow}

\def\lf{\left}
\def\rg{\right}

\begin{document}


\title{Parity--violating DIS and the flavour dependence of the EMC effect}

\author{I.~C.~Clo\"et}
\affiliation{CSSM and ARC Centre of Excellence for Particle Physics at the Terascale, \\ 
School of Chemistry and Physics, 
University of Adelaide, Adelaide SA 5005, Australia
}

\author{W.~Bentz}
\affiliation{Department of Physics, School of Science, Tokai University,
             Hiratsuka-shi, Kanagawa 259-1292, Japan}

\author{A.~W.~Thomas}
\affiliation{CSSM and ARC Centre of Excellence for Particle Physics at the Terascale, \\ 
School of Chemistry and Physics, 
University of Adelaide, Adelaide SA 5005, Australia
}

\begin{abstract}
Isospin-dependent nuclear forces play a fundamental role in nuclear structure. 
In relativistic models of nuclear structure constructed at the quark level 
these isovector nuclear forces affect the $u$ and $d$ quarks differently, leading to
non-trivial flavour dependent modifications of the nuclear parton distributions. 
We explore the effect of isospin dependent forces for parity-violating deep 
inelastic scattering on nuclear targets and demonstrate that 
the cross-sections for nuclei with $N \neq Z$ are sensitive to  
the flavour dependence of the EMC effect.
Indeed, for nuclei like lead and gold 
we find that 
these flavour dependent effects are large.
\end{abstract}


\maketitle


Understanding the mechanisms responsible for the change in the per-nucleon deep 
inelastic scattering (DIS) cross-section between the deuteron and heavier nuclei remains 
one of the most important challenges confronting the nuclear physics community.
In the valence quark region this effect is characterized by 
a quenching of the nuclear structure functions relative to 
those of the free nucleon and is known as the EMC effect~\cite{Aubert:1983xm}.
This discovery has led to a tremendous amount of experimental and theoretical 
investigation~\cite{Geesaman:1995yd,Berger:1987er,Frankfurt:1988nt}.  
However, after the passage of more than 25 years there remains no broad consensus 
regarding the underlying mechanism responsible for the EMC effect.

Early attempts to explain the EMC effect focused on detailed nuclear structure
investigations \cite{Bickerstaff:1989ch} and the possibility of an enhancement in the pionic component of 
the nucleon in-medium~\cite{Llewellyn Smith:1983qa,Ericson:1983um}. The former studies
were unable to describe the data and the latter explanation appears to be ruled out
by Drell-Yan measurements of the anti-quark distributions in nuclei~\cite{Alde:1990im}.
Other ideas included the possibility of exotic,  
six-quark bags in the nucleus~\cite{Jaffe:1982rr} or 
traditional short-range correlations~\cite{Weinstein:2010rt}.
It has also been argued that the EMC effect
is a result of changes in the internal structure of 
the bound nucleons brought about
by the strong nuclear fields inside the nucleus~\cite{Saito:1992rm}. 
Many of these approaches can explain the qualitative 
features of the EMC effect but the underlying physics mechanisms differ 
substantially.  

To make further progress in our understanding of the mechanism responsible for 
the EMC effect, it has become clear that we require new experiments that reveal 
genuinely novel features of this effect.
In this Letter we propose an important step in this direction, namely the exploitation of 
parity-violating DIS (PVDIS), which follows 
from the interference between photon and $Z^0$ exchange. 
When used in conjunction with the familiar electromagnetic DIS data, 
it becomes possible to obtain explicit information about
the quark flavour dependence of the nuclear parton 
distribution functions (PDFs). 
This will allow the predictions of any model of the 
EMC effect to confront new experimental information and hence 
provide important insights into this longstanding puzzle.

The parity violating effect of the interference between photon and $Z^0$ exchange,
in the DIS of longitudinally polarized electrons on an unpolarized target, 
leads to the non-zero asymmetry
\begin{align}
A_{PV} = \frac{\s_R - \s_L}{\s_R + \s_L},
\end{align}
where $\s_L$ and $\s_R$ denote the double differential cross-sections for DIS 
of right- and left-handed polarized electrons, respectively.
In the Bjorken limit $A_{PV}$ can be expressed as~\cite{Brady:2011uy}
\begin{align}
A_{PV} = \frac{G_F\,Q^2}{4\sqrt{2}\,\pi\,\a_{\text{em}}}
\lf[a_2(x_A) + \frac{1-(1-y)^2}{1+(1-y)^2}\,a_3(x_A)\rg],
\label{eq:APV_Bjorken}
\end{align}
where $x_A$ is the Bjorken scaling variable of 
the nucleus multiplied by $A$, 
$G_F$ is the Fermi coupling constant and $y=\nu/E$ 
is the energy transfer divided by the incident electron energy. 
The $a_2$ term in Eq.~\eqref{eq:APV_Bjorken} originates from the product
of the electron weak axial current and the quark weak vector 
current and has the form
\begin{align}
a_2(x_A) = -2\,g_A^e\,\frac{F_{2A}^{\g Z}(x_A)}{F_{2A}^\g(x_A)} 
= \frac{2\sum_q e_q\,g_V^q\,q_A^+(x_A)}{\sum_q e_q^2\,\,q_A^+(x_A)}.
\label{eq:a2}
\end{align}
The plus-type quark distributions are defined by 
$q_A^+(x_A) = q_A(x_A) + \bar{q}_A(x_A)$, $e_q$ is the quark charge,
$g_A^e = -\tfrac{1}{2}$~\cite{Amsler:2008zzb} and the quark weak vector couplings are~\cite{Amsler:2008zzb}
\begin{align}
\hs{-2mm} g_V^u = \frac{1}{2} - \frac{4}{3}\,\sin^2\!\theta_W, \hs{5mm}
          g_V^d = -\frac{1}{2} + \frac{2}{3}\,\sin^2\!\theta_W, 
\end{align}
where $\theta_W$ is the weak mixing angle. 
The parity violating $F_2$ structure function 
of the target arising from $\g Z$ interference 
is labelled as $F_{2A}^{\g Z}(x_A)$, 
while $F_{2A}^\g(x_A)$ is the familiar electromagnetic 
structure function of traditional DIS.
The parton model expressions for these structure functions are~\cite{Amsler:2008zzb}
\begin{align}
F_{2A}^{\g Z} = 2\,x_A\sideset{}{_q}\sum e_q\,g_V^q\,  q_A^+, \hs{3mm}
F_{2A}^\g    =    x_A\sideset{}{_q}\sum e_q^2\,      q_A^+.
\end{align}
The $a_3$ term in Eq.~\eqref{eq:APV_Bjorken} is given by 
\begin{align}
%
a_3(x_A) = -2\,g_V^e\,\frac{F_{3A}^{\g Z}(x_A)}{F_{2A}^\g(x_A)}
= -4\,g_V^e\,\frac{\sum_q e_q\,g_A^q\,q_A^-(x_A)}{\sum_q e_q^2\,\,q_A^+(x_A)}.
\end{align}
where $g_V^e = -\tfrac{1}{2} + 2\sin^2\theta_W$,
$g_A^u = -g_A^d = \tfrac{1}{2}$~\cite{Amsler:2008zzb} 
and $q_A^-(x_A) = q_A(x_A) - \bar{q}_A(x_A)$.
This term is suppressed in the parity-violating asymmetry, $A_{PV}$,
because of its $y$-dependent prefactor and the fact that $g_V^e \ll g_A^e$.
Therefore, the $a_3$ term will not be considered any further in this Letter.

The parity violating structure function, $F^{\g Z}_{2A}$, has a different flavour 
structure from that of $F^{\g}_{2A}$ and, as a consequence, $a_2(x)$ is
sensitive to flavour dependent effects. 
To illustrate this we expand $a_2$ about the $u_A \simeq d_A$ limit, 
by ignoring heavy quark flavours we obtain
\begin{align}
a_2(x_A) \simeq \frac{9}{5} - 4\sin^2\theta_W
- \frac{12}{25}\,
\frac{u^+_A\lf(x_A\rg) - d^+_A\lf(x_A\rg)}{u^+_A\lf(x_A\rg) + d^+_A\lf(x_A\rg)}.
\label{eq:a2isovector}
\end{align}
A measurement of $a_2(x_A)$ will therefore provide information about the 
flavour dependence of the nuclear quark 
distributions and when coupled with existing measurements of $F^{\g}_{2A}$ a
reliable extraction of the flavour dependent quark distributions
becomes possible.
Alternatively, if the isovector correction term in Eq.~\eqref{eq:a2isovector}
is known, then the parity violating asymmetry provides an independent method 
with which to determine the weak mixing angle. For example, if we 
ignore heavy quark flavours,
quark mass differences~\cite{Londergan:2009kj,Rodionov:1994cg,Sather:1991je} and
electroweak corrections, the $u$- and $d$-quark distributions of an isoscalar 
target will be identical, and in this limit Eq.~\eqref{eq:a2isovector} becomes
\begin{align}
a_2(x_A) \stackrel{N=Z}{\lra} \frac{9}{5} - 4\sin^2\theta_W.
\label{eq:a2_isoscalar}
\end{align}
This result is analogous to the Paschos-Wolfenstein 
ratio~\cite{Paschos:1972kj,Cloet:2009qs}
in neutrino DIS, which motivated the NuTeV collaboration measurement of 
$\sin^2\theta_W$~\cite{Zeller:2001hh,Bentz:2009yy}. 
An important advantage of $a_2(x_A)$ as a measure of the weak mixing angle
is that in the valence quark region strange quark effects are almost
absent, which eliminates the largest uncertainty in the NuTeV measurement 
of $\sin^2\theta_W$~\cite{Bentz:2009yy}. Also the isovector correction term in 
Eq.~\eqref{eq:a2isovector} does not depend on $\sin^2\theta_W$ and thus a 
measurement of $a_2(x_A)$ at each value of $x_A$ constitutes a
separate determination of $\sin^2\theta_W$.
More importantly however, in the context of this work, is that $a_2$ is 
sensitive to flavour dependent nuclear effects that influence 
the quark distributions 
of nuclei. Indeed, because of this sensitivity,  
a measurement of $a_2$ on a target with $N > Z$  would 
provide an excellent opportunity to test the 
importance of the isovector 
EMC effect~\cite{Cloet:2009qs,Bentz:2009yy} for the interpretation of  
the anomalous NuTeV result for $\sin^2\theta_W$.

To determine the nuclear quark distributions and investigate their isospin 
dependence we use the Nambu--Jona-Lasinio (NJL) model~\cite{Nambu:1961tp,Nambu:1961fr}, 
which is a QCD motivated low energy chiral effective theory characterized by a 4-fermion 
contact interaction between the quarks.
The NJL model has a long history of success in describing mesons as $\bar{q}q$ bound 
states~\cite{Vogl:1991qt,Hatsuda:1994pi}
and more recently as a self-consistent model for free and in-medium 
baryons~\cite{Cloet:2006bq,Cloet:2005rt,Cloet:2005pp,Mineo:2003vc}. 

The NJL interaction Lagrangian can be decomposed into various 
$\bar{q}q$ and $qq$ interaction channels via Fierz transformations~\cite{Ishii:1995bu},
where relevant details to this discussion are given in Ref.~\cite{Cloet:2006bq}.
The scalar $\bar{q}q$ interaction dynamically generates a constituent quark 
mass via the gap equation and gives rise to an isoscalar-scalar mean field in-medium.
The vector $\bar{q}q$ interaction terms are used to generate the isoscalar-vector, 
$\omega_0$, and isovector-vector, $\rho_0$, mean-fields in-medium. 
The $qq$ interaction terms give the diquark $t$-matrices
with poles corresponding to the scalar and axial-vector diquark masses. 
The nucleon vertex function and mass are obtained by solving the homogeneous 
Faddeev equation for a quark and a diquark, 
where the static approximation is used to truncate the
quark exchange kernel~\cite{Cloet:2005pp}. 
To regulate the NJL model we choose the
proper-time scheme, which eliminates unphysical thresholds for nucleon decay 
into quarks, and hence simulates an important aspect of QCD, 
namely quark confinement~\cite{Ebert:1996vx,Hellstern:1997nv,Bentz:2001vc}.

To self-consistently determine the strength of the mean scalar and vector fields,
an equation of state for nuclear matter is derived from the NJL Lagrangian
using hadronization techniques~\cite{Bentz:2001vc}. In a mean-field approximation
the result for the energy density is~\cite{Bentz:2001vc}
\begin{equation}
\mathcal{E} = \mathcal{E}_V - \frac{\o_0^2}{4\,G_\o} - 
\frac{\rho_0^2}{4\,G_\rho} + \mathcal{E}_p + \mathcal{E}_n,
\label{eq:effective}
\end{equation}
where $G_\o$ and $G_\rho$ are the $\bar{q}q$ couplings in 
the isoscalar-vector and isovector-vector
channels respectively. The vacuum energy, $\mathcal{E}_V$, 
has the familiar Mexican hat shape and the
energies of the protons and neutrons  moving through the 
mean scalar and vector fields are labelled
by $\mathcal{E}_p$ and $\mathcal{E}_n$, respectively. 
Minimizing the effective potential 
with respect to each vector field gives the following relations:
$\omega_0 = 6\,G_\omega\lf(\rho_p + \rho_n\rg)$ 
and $\rho_0   = 2\,G_\rho\lf(\rho_p - \rho_n\rg)$,
where $\rho_p$ is the proton and $\rho_n$ the neutron density.
The vector field experienced by each quark flavour is given by
$V_u = \omega_0 + \rho_0$ and $V_d = \omega_0 - \rho_0$.

\begin{figure}[t]
\subfloat{\centering\includegraphics[width=\columnwidth,clip=true,angle=0]{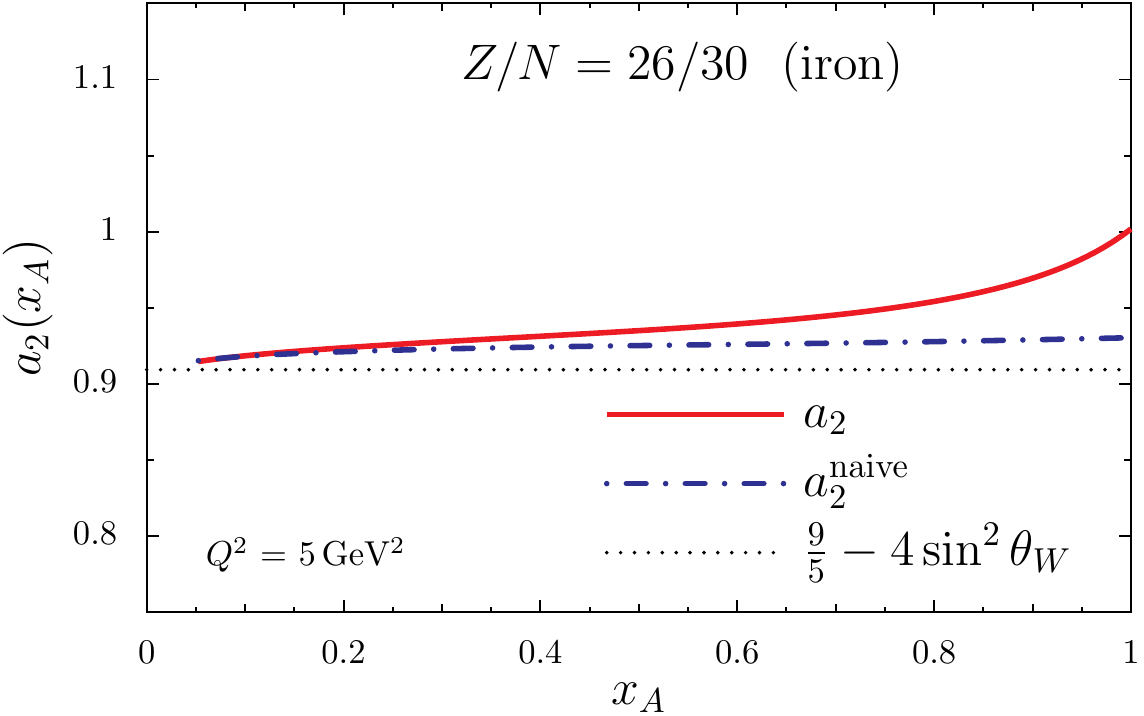}}\\
\subfloat{\centering\includegraphics[width=\columnwidth,clip=true,angle=0]{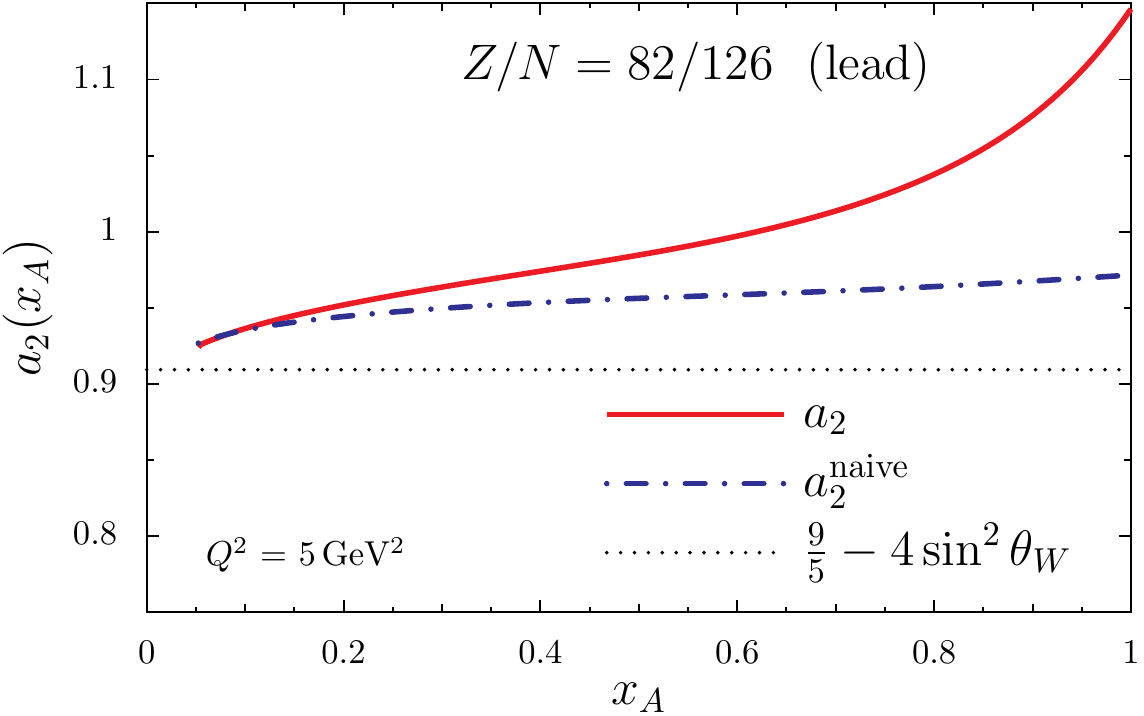}}
\caption{Asymmetric nuclear matter results for $a_2(x_A)$ obtained by using the $Z/N$ ratio of iron (top) and
lead (bottom). In each figure the dotted line is the isoscalar result, the dot-dashed line 
the naive expectation where no medium effects have been included and the solid line is the 
full result.}
\label{fig:a2}
\end{figure}

The parameters of the model are determined by reproducing standard hadronic properties, 
such as masses and decay constants, 
as well as the empirical saturation energy and density of symmetric nuclear matter. 
The empirical symmetry energy of nuclear matter, namely $a_4=32\,$MeV, is used to 
constrain $G_\rho$, giving $G_\rho = 14.2\,$GeV$^{-2}$. A discussion of the model
parameters can be found in Ref.~\cite{Cloet:2006bq}.

Using medium modified quark distributions -- calculated following the techniques 
of Refs.~\cite{Mineo:2003vc,Cloet:2005rt} -- we determine $a_2(x_A)$ for symmetric and 
asymmetric nuclear matter. In each case the total baryon density, 
$\rho_B = \rho_p + \rho_n$, is kept fixed and only the proton-neutron ratio is varied.
In Figs.~\ref{fig:a2} we present
our results for $a_2(x_A)$ in nuclear matter with a proton-neutron ratio equal to that 
of iron (top) and lead (bottom). The full result, which
includes the effects from Fermi motion and the scalar and vector mean-fields, is 
represented by the solid line. The dot-dashed line is the naive expectation where the 
nuclear quark distributions are obtained from the free proton and neutron PDFs 
without modification. The dotted line is the result for isoscalar nuclear matter,
which is given by Eq.~\eqref{eq:a2_isoscalar}
and maybe a reasonable approximation to nuclei such as $^{12}$C and $^{40}$Ca.
When evaluating $g_V^q$ we have used the on-shell renormalization scheme value 
of $\sin^2\theta_W = 0.2227$~\cite{Zeller:2001hh}. 

The leading correction to $a_2$ is isovector, as illustrated in Eq.~\eqref{eq:a2isovector}. 
As a consequence, the difference between the naive
and full results of Figs.~\ref{fig:a2} is primarily 
caused by the non-zero $\rho^0$ mean-field. 
This is precisely the same effect which eliminates 1 to  
1.5$\sigma$~\cite{Cloet:2009qs,Bentz:2009yy} of the NuTeV discrepancy  
with respect to the Standard Model in their measurement
of $\sin^2 \theta_W$. Thus, quite apart from the intrinsic importance of 
understanding the dynamics of quarks within nuclei, 
the observation of these large flavour dependent nuclear effects illustrated
in Figs.~\ref{fig:a2} would be direct evidence that
the isovector EMC effect plays an important role in interpreting the
NuTeV data. It would also indicate the importance of flavour dependent effects 
in our understanding of the EMC effect in nuclei like lead and gold,
a point we will return to shortly.

The $a_2$ function is potentially sensitive to charge symmetry violation (CSV) effects aswell,
which are a consequence of the light quark mass differences and electroweak 
corrections~\cite{Londergan:2009kj,Rodionov:1994cg,Sather:1991je}. 
Including only the CSV correction, Eq.~\eqref{eq:a2isovector} becomes
\begin{align}
a_2(x) \simeq \frac{9}{5} - 4\sin^2\theta_W
- \frac{6}{25}\,
\frac{\delta u^+(x) - \delta d^+(x)}{u^+_p(x) + d^+_p(x)},
\label{eq:a2CSV}
\end{align}
where $\delta u^+ \equiv u_p^+ - d_n^+$ and  $\delta d^+ \equiv d_p^+ - u_n^+$.
These effects are largely independent of the in-medium effects already 
discussed~\cite{Bentz:2009yy} and using the 
MRST parametrizations of Ref.~\cite{Martin:2003sk}, with their central value of $\kappa = -0.2$, 
we find this correction to be negligible on the scale of Figs.~\ref{fig:a2}.
However, if these CSV effects turn out to be larger than expected, they can be
constrained via measurements on isospin symmetric nuclei, where the isovector EMC 
corrections are mucj smaller.

\begin{figure}[t]
\subfloat{\centering\includegraphics[width=\columnwidth,clip=true,angle=0]{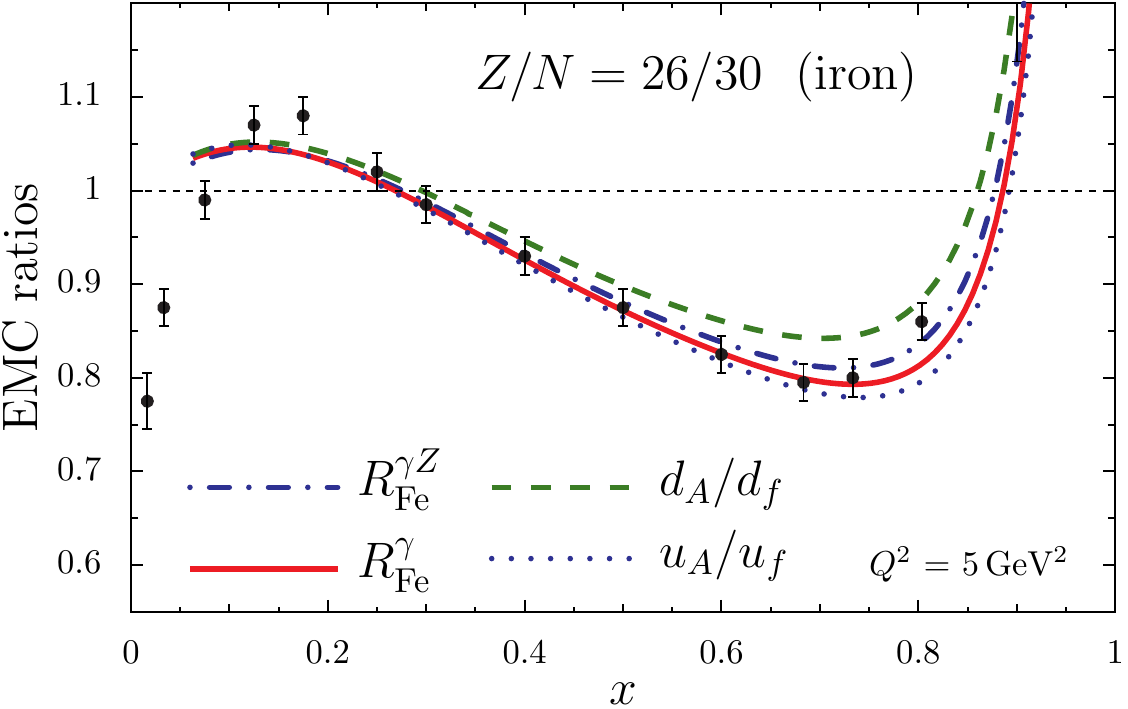}} \\
\subfloat{\centering\includegraphics[width=\columnwidth,clip=true,angle=0]{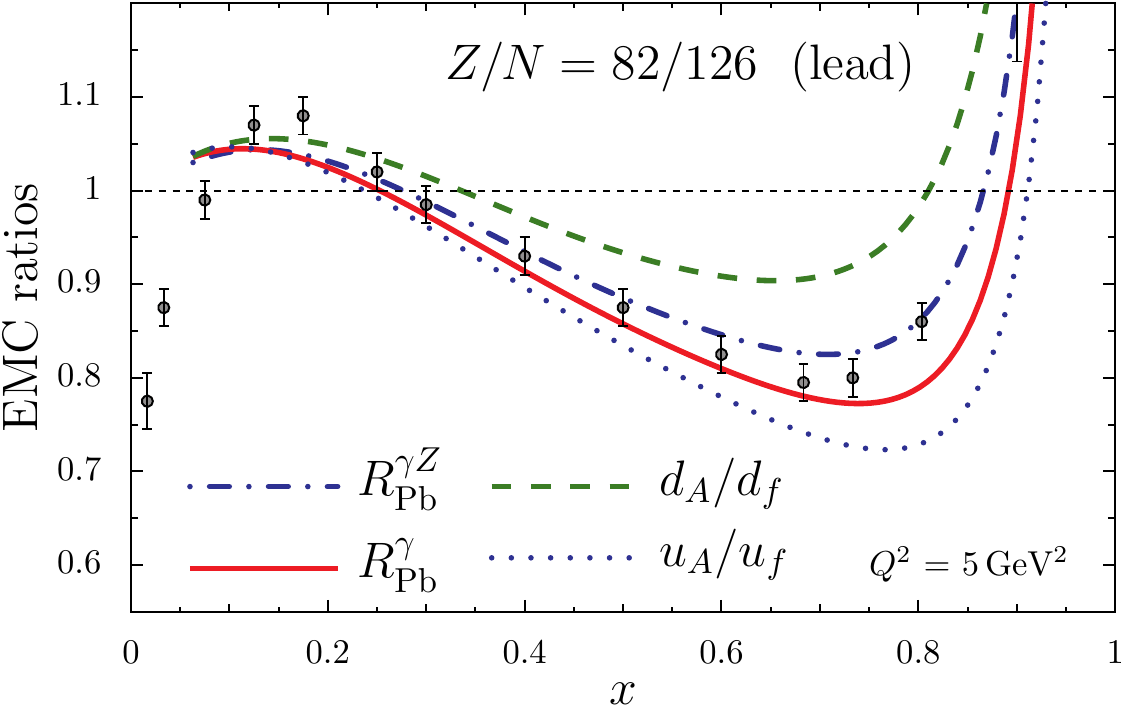}}
\caption{The solid line in each figure
is our full result for the EMC effect in electromagnetic DIS on nuclear matter, with the 
$Z/N$ ratio chosen to be that of iron (top) and lead (bottom). The dash-dotted line
illustrates the EMC ratio for PVDIS on the same target. 
The dotted and dashed lines show the EMC effect in the $u$ and $d$ quark
sectors, respectively. The data in each figure is for isoscalar
nuclear matter and is taken from Ref.~\cite{Sick:1992pw}. }
\label{fig:EMC}
\end{figure}

The EMC effect can be defined for both electromagnetic and parity violating DIS
via the ratio
\begin{align}
R^i = \frac{F_{2A}^i}{F_{2A,\text{naive}}^i} = \frac{F_{2A}^i}{Z\,F_{2p}^i + N\,F_{2n}^i},
\label{eq:EMCeffect}
\end{align}
where $i \in \g,\,\g Z$. The corresponding proton and neutron 
structure functions are respectively labelled by $F_{2p}^i$ and $F_{2n}^i$, 
while $F_{2A}^i$ is the structure function of the target. The naive structure
function $F_{2A,\text{naive}}^i$ has no medium effects whatsoever, and therefore,
in this limit $R^i$ would be unity. Expressing the EMC effect in terms of 
the quark distributions we find the parton model expressions
\begin{align}
R^\gamma    \simeq \frac{4\,u^+_A + d^+_A}{4\,u^+_f + d^+_f}, \hs{10mm}
R^{\gamma Z} \simeq \frac{1.16\,u^+_A + d^+_A}{1.16\,u^+_f + d^+_f},
\end{align}
where $q_f$ are the quark distributions of the target if it were composed of free nucleons.
For an isoscalar target we have $R^\gamma = R^{\gamma Z}$
(modulo electroweak, quark mass and heavy quark flavour effects).
However, for nuclei with $N \neq Z$ these two EMC effects need not be equal. 
The solid line in Figs.~\ref{fig:EMC} illustrates our
EMC effect results for $F_{2A}^\g$ in nuclear matter,
with $Z/N$ ratios equal to that of iron (top) and lead (bottom), while the 
corresponding EMC effect in $F_{2A}^{\g Z}$ is represented by the dot-dashed line. 
We find that as the proton-neutron ratio is decreased, the EMC effect in $F_{2A}^\g$
increases, whereas the EMC effect in $F_{2A}^{\g Z}$ is slightly reduced.
Consequently, for $N > Z$ nuclei we find that $R^\gamma < R^{\gamma Z}$ on the domain 
$x_A \gtrsim 0.2$, which is the domain over which our 
valence quark model can be considered reliable.

The fact that $u_A/u_f < d_A/d_f$ and as a consequence $R^\gamma < R^{\gamma Z}$ in nuclei with 
a neutron excess is a direct
consequence of the isovector mean field and is a largely model independent 
result. In Ref.~\cite{Cloet:2009qs} in was demonstrated that the isovector mean field leads to
a small shift in quark momentum from the $u$ to the $d$ quarks, and hence, the
in-medium depletion of $u_A$ is stronger than that of $d_A$ in the valence
quark region. Because $u_A$ is multiplied by a factor four in the ratio $R^\gamma$, the
depletion is more pronounced for this ratio than for $R^{\gamma Z}$, where the $d$-quark
quickly dominates as $Z/N$ becomes less than one. The results presented in Figs.~\ref{fig:EMC} 
demonstrate that the flavour dependence of the EMC effect is potentially large in nuclei like lead 
and gold.

We have shown that an accurate comparison of the electromagnetic and parity 
violating DIS cross sections have the potential to pin down the flavour 
dependence of the EMC effect. The most direct determination of this flavour 
dependence (c.f. Figs.~\ref{fig:EMC}) would involve charged current reactions on 
heavy nuclei at an electron-ion collider~\cite{Thomas:2009ei} or with certain 
Drell-Yan reactions~\cite{Dutta:2010pg,Chang:2011ra}.
However, such experiments will not be possible for ten to twenty years. 
On the other hand, accurate measurements of PVDIS on heavy nuclei should be possible 
at Jefferson Lab after the 12 GeV upgrade~\cite{pvdisloi} and would therefore 
provide a timely, critical test of an important class of models which aim to describe the modification 
of the nuclear structure functions. These experiments would complement alternative 
methods to access the quark substructure of nuclei, for example, the measurement of the EMC effect for spin 
structure functions~\cite{Cloet:2005rt,Cloet:2006bq}, and as a corollary, would also 
offer a unique insight into the description of nuclear structure 
at the quark level. Finally, they would constitute a direct test of 
the isovector EMC effect correction to the NuTeV measurement of $\sin^2\theta_W$.

\section*{Acknowledgements}
\vspace*{-5mm}
The work is supported by the ARC Centre of Excellence in Particle Physics at the Terascale 
and an Australian Laureate Fellowship FL0992247 (AWT).

\end{document}